# A plausible link between the asteroid 21 Lutetia and CH carbonaceous chondrites

Carles E. MOYANO-CAMBERO[1*], Josep M. TRIGO-RODRIGUEZ[1], Jordi LLORCA[2], Sonia FORNASIER[3], Maria A. BARUCCI[3] and Albert RIMOLA[4]

[1]Institut de Ciències de l'Espai (CSIC-IEEC), Campus UAB, Carrer de Can Magrans, s/n, 08193 Cerdanyola del Vallès (Barcelona), Spain
[2]Institut de Tècniques Energètiques i Centre de Recerca en Nanoenginyeria, Universitat Politècnica de Catalunya, Diagonal 647, ETSEIB, 08028 Barcelona, Spain
[3]LESIA, Observatoire de Paris, PSL Research University, CNRS, Univ. Paris Diderot, Sorbonne Paris Cité, UPMC Univ. Paris 06, Sorbonne Universités, 5 Place J. Janssen, 92195 Meudon Principal Cedex, France
[4]Departament de Química, Universitat Autònoma de Barcelona, 08193 Bellaterra, Spain;
*Corresponding author. E-mails: moyano@ice.csic.es ; trigo@ice.csic.es

**Abstract** - A crucial topic in planetology research is establishing links between primitive meteorites and their parent asteroids. In this study we investigate the feasibility of a connection between asteroids similar to 21 Lutetia, encountered by the Rosetta mission in July 2010, and the CH3 carbonaceous chondrite Pecora Escarpment 91467 (PCA 91467). Several spectra of this meteorite were acquired in the ultraviolet to near-infrared (0.3 to 2.2 µm) and in the mid-infrared to thermal infrared (2.5 to 30.0 µm or 4000 to ~333 cm$^{-1}$), and they are compared here to spectra from the asteroid 21 Lutetia. There are several similarities in absorption bands and overall spectral behavior between this CH3 meteorite and 21 Lutetia. Considering also that the bulk density of Lutetia is similar to that of CH chondrites, we suggest that this asteroid could be similar, or related to, the parent body of these meteorites, if not the parent body itself. However, the apparent surface diversity of Lutetia pointed out in previous studies indicates that it could simultaneously be related to other types of chondrites. Future discovery of additional unweathered CH chondrites could provide deeper insight in the possible connection between this family of metal-rich carbonaceous chondrites and 21 Lutetia or other featureless, possibly hydrated high-albedo asteroids.

**Tables: 2**
**Figures: 7**

# INTRODUCTION

Establishing relationships between meteorites and their possible parent asteroids has always been quite a tricky task, as there are many differences in the observing conditions, resolutions and methods used for the study of these primordial objects (see e.g. Clark et al. 2010, 2011). To complicate the overall puzzle it is obvious that the physical processes at work in each one of the parent bodies of meteorites have been remarkably different depending on their particular evolutionary stories (Binzel et al. 2010). In this paper we present evidence about the similitude, and even plausible connection, between almost featureless, possibly hydrated high-albedo asteroids, and one group of carbonaceous chondrites (hereafter CCs), characterized by a particularly high content in metal and a very low degree of alteration: the metal-rich CHs. Few asteroids fit this description: 21 Lutetia, 77 Frigga, 135 Hertha, 136 Austria, and possibly others (Hardersen et al. 2011). Since 21 Lutetia (or simply Lutetia), encountered by the Rosetta mission in July 2010, has been studied in greater detail than the others, we compare this asteroid to the CH chondrites on the basis of spectral similarities in the ultraviolet to thermal infrared (UV-IR) range, and considering our current knowledge on both the asteroid and this type of meteorites.

Lutetia is a large main-belt asteroid. Some of its spectra are flat and almost featureless in the 0.15 to 2.5 µm range (Birlan et al. 2004; Ockert-Bell et al. 2010; Coradini et al. 2011; Hardersen et al. 2011; Sierks et al. 2011), but others can show faint signatures, depending mostly on the region sampled (Birlan et al. 2006; Weaver et al. 2010; Stern et al. 2011; Barucci et al. 2012), which makes the spectral type classification of this asteroid quite difficult. Tentatively, and due to its moderately-high albedo and presumably high metal content, it has been considered as an M taxonomic type asteroid into the X-group according to the Tholen classification (Barucci et al. 1987; Tholen 1989). In the SMASS classification, it has been described as an uncommon Xk transition type asteroids into the X-group or one of the very rare Xc transition type asteroids, due to the its spectral slope and absorption bands (Bus and Binzel, 2002; Ockert-Bell et al. 2008; DeMeo et al. 2009; Shepard et al. 2010). Both its albedo and bulk density are too low to be considered a largely metallic asteroid (Weaver et al. 2010), but they are also too high for the typical C-class asteroids (Shepard et al. 2008), historically considered as the parent bodies of most CCs (see e.g. Chapman et al. 1975). Indeed, its albedo seems to be more consistent with enstatite chondrites, CH chondrites, or silicate bearing iron meteorites (Shepard et al. 2008).

In addition to spectral shape comparison, we analyze the similarities of the CH group of CCs and Lutetia on the basis of the mineralogy of these meteorites and their relatively high reflectance, compared to most CCs, which is probably associated with their high abundance of 10 to 200 µm metal grains (Campbell and Humayun, 2004; Trigo-Rodríguez et al. 2014). Indeed, the albedo of asteroid Lutetia is also higher than expected for the kind of dark asteroids commonly associated with CCs. Besides, the spectra obtained from several regions of this asteroid, which usually show few features, are consistent with the commonly flat spectra characteristic of most CCs, pointing towards a closer relation with these families of meteorites than with other types of chondrites. However, we take into account that some of the spectra associated to Lutetia show features consistent with other meteorites, which could imply that, even if CH meteorites are indeed connected to asteroids similar to Lutetia, they could not be representative of the entire surface of their parent body.

## ANALYTICAL PROCEDURE

**Rationale for sample selection**

Several attempts to establish Lutetia as the parent body of different meteorites groups have been done before. Primitive meteorites (of petrologic type 3 or close) have been related to this body due to observations in the UV-NIR range (e.g. Nedelcu et al. 2007, where meteorite and asteroid spectra are compared using a $\chi^2$ fitting test). Carbonaceous bodies also seem to match some of the spectral properties of Lutetia, and indeed, it has been related to COs and CVs due to its mid-to-far-infrared features and polarimetry data (Lazzarin et al. 2004, 2009; Barucci et al. 2008, 2012; Belskaya et al. 2010). However, most spectra of Lutetia do not show the lack of the drop-off below 0.55 µm which is common in CVs (Gaffey 1976)., and both groups of CCs show the 1 µm band related to the presence of olivine (Vernazza et al. 2011). Despite the relatively high abundance of mostly featureless (in the UV-NIR range) matrix on those meteorites, this band is still too strong considering that it has not been reported on Lutetia. Actually, rather than making those features disappear the presence of matrix material darkens the spectra of these meteorites, which makes them less reflective than expected for a meteorite whose parent asteroid is Lutetia (Drummond et al. 2010). This asteroid has also been compared with meteorites with a very high content in metal, due to its moderately high albedo around 0.2 (see e.g. Mueller et al. 2006, for a summary of albedo determinations for Lutetia in table 5, and Sierks et al. 2011, for more recent result obtained with the cameras onboard Rosetta), but the inferred density of this asteroid, ~ 3400 kg/m$^3$ (Weiss et al. 2010), is too small for an object dominated by metallic components, even assuming a typical porosity of 10-15% (Weaver et al. 2010; Pätzold et al. 2011). It has been suggested that enstatite chondrites (ECs) match the spectra of 21 Lutetia better than most CCs (see, e.g., Ockert-Bell et al. 2010; Vernazza et al. 2011). As their name indicates, enstatite chondrites have FeO-free enstatite as their main forming mineral (Weisberg et al. 2006), which is almost featureless in the UV-NIR range (see e.g. Klima et al. 2007; Vernazza et al. 2011). Furthermore, a few Lutetia spectra show a weak band at ~0.9 µm (see e.g. Birlan et al. 2006) that could be indicating the presence of enstatite on the asteroid's surface (Klima et al. 2007; Vernazza et al. 2011). Lutetia spectra in the mid-infrared apparently implies that the silicates on its surface are most likely iron free, like enstatite (Vernazza et al. 2011), and that this asteroid has not suffered from aqueous alteration (Coradini et al. 2011, Vernazza et al. 2011), consistently with the 3 to 6 petrographic type of enstatite chondrites (Weisberg et al. 2006). However, other studies suggest that the far-infrared Spitzer data of Lutetia does not match that of enstatite chondrites (Barucci et al. 2008; Perna et al. 2010), and that these meteorites neither show the Christiansen peak around 9.3 µm (~1075 cm$^{-1}$), reported in a spectrum of 21 Lutetia and common in CCs (Izawa el al. 2010; Barucci et al. 2012). Additionally, enstatite chondrites do not show the features in the visible related to aqueous alteration and hydrated minerals that have been found in some Lutetia spectra (Lazzarin et al. 2009; Hardersen et al. 2011; Rivkin et al. 2011).

Some authors (e.g. Lazzarin et al. 2009; Shepard et al. 2010; Coradini et al.; 2011) took into account the metal-rich groups of CCs (CRs, CHs and CBs) in the analysis of Lutetia spectra. It has been shown that the densities of these meteorites are consistent with that of the 21 Lutetia asteroid (Coradini et al. 2011), specifically in the case of CHs, with a bulk density of ~ 3650 kg/m$^3$ (Macke et al. 2010). The geometric albedo of this asteroid, determined precisely in situ by the Optical, Spectroscopic, and Infrared Remote Imaging System (OSIRIS) instrument onboard Rosetta, has a value of 0.19 ± 0.01 (Sierks et al. 2011). This value corresponds to zero phase angle, and therefore in previous studies it has been corrected using Lutetia phase function, in order to be compared to albedos of meteorites (usually measured at angles of ~7 to 10 degrees), obtaining a value in the range of 0.13-0.16 (Belskaya et al. 2010). It is larger than the usual values for other primitive bodies and most CCs, and lower than the ones from metal meteorites (Gaffey, 1976) but in agreement with metal-rich CCs (Weiss et al. 2012, Trigo-Rodríguez et al. 2014). However, the particle size can have a strong effect on albedo, so the fact that Lutetia is probably covered in fine-grained regolith has to be taken into account to do a proper comparison with the spectral properties of meteorites (Belskaya et al. 2010). Moreover, Lutetia shows surface heterogeneities (see Barucci et al. 2012, and references therein), indicating the presence of different lithologies and therefore the possibility that more than one type of meteorites could have their origin in such a parent body (Nedelcu et al. 2007). We propose, therefore, that these metal-rich CCs should be considered as possible analogues of certain regions of this or similar asteroids, which does not mean that the whole body shows this kind of metal-rich carbonaceous composition.

We know from several previous studies that each chondrite group shows distinctive reflection spectra (see e.g., Cloutis et al. 2012a; 2012b, where they use spectra from powders of different sizes, and Trigo-Rodríguez et al. 2014, where we mostly obtained spectra from meteorite thin or thick sections). CCs belonging to the CR, CH and CB groups show rather high reflection degrees, increasing from CRs to CBs. This is probably due, at least partially, to their high content in metal grains. However this relationship is not linear as the distribution of metal in these three groups is heterogeneous: while CH chondrites have most of their metal in the matrix, CRs show metal inclusions mostly inside the chondrules, which would not necessary imply a high increase of the albedo of their parent body (Krot et al. 2002; Trigo-Rodríguez et al. 2014). On the other hand, CBs have very high metal abundances mostly out of the chondrules, so their parent body should show a particularly high albedo in the same way that the meteorite samples show a very high reflectance spectrum (Trigo-Rodríguez et al. 2014). The 0.13-0.16 reflectance of Lutetia at phase angles between 7 and 10 degrees is low compared to iron meteorites and CBs, and high compared to other CCs, including CRs, but it is comparable to the absolute reflectance obtained for a CH chondrite in a previous study, between 0.10 and 0.20 in the 0.2 to 2.0 µm range. (see e.g. Fig. 1d of our Trigo-Rodríguez et al. 2014, paper). This is why we consider that among metal rich CCs, CH chondrites are the more plausible candidates to be related to Lutetia or similar asteroids.

CH chondrites are composed of ~70% of small cryptocrystalline chondrules (20-90 µm in diameter), a relatively high (~20%) amount of FeNi-metal grains with approximately solar Ni/Co ratio, and hydrated fine-grained matrix-like clasts instead of matrix, which include Fe-rich phyllosilicates (such as serpentine), magnetites, sulfides and carbonates (Greshake et al. 2002; Krot et al. 2002; Weisberg et al. 2006). Those meteorites are highly depleted in volatiles and moderately-volatile lithophile elements (Weisberg et al. 2006), and show a low amount of Calcium-Aluminum Inclusions (CAIs) (Weisberg et al. 2006; Cloutis et al. 2012b). Similarly to enstatite chondrites, Mg-rich pyroxenes are the main silicate on CH chondrites, and the existing olivines are also Mg-rich (Weisberg et al. 2006). Two formative scenarios are suggested for those meteorites: a subchondritic origin as a result of asteroidal collision, or formation as pristine products of the solar nebula (Weisberg et al. 2006, and references therein).

For this study we obtained some samples from the high-metal carbonaceous chondrite of petrologic type 3 (CH3) Pecora Escarpment (PCA) 91467, recovered by the Antarctic Search for Meteorites (ANSMET) program in 1991, and nowadays belonging to the NASA Antarctic collection. These samples consist of a small chip, the thin section PCA 91467,25 (Fig. 1), and the thicker section PCA 91467,16, from which a second thin section was obtained. Although a B/C weathering grade of B/C has been indicated for this meteorite (Bischoff et al. 1994), these samples show a remarkably pristine interior that can be representative of the forming materials of its progenitor asteroid, if we deal appropriately with the differences in scale and other factors. This particular CH chondrite is consistent with the other members of its type, with abundant lithic fragments and few relatively small chondrules, Mg-rich pyroxenes and olivines with mostly $Fs_{1-5}$ and $Fa_{1-4}$ composition, and abundant NiFe metal and hydrated materials, such as magnetite (Bischoff et al. 1994; Sugiura 2000; Cloutis et al. 2012b).

**Spectroscopy in the 0.3 to 2.2 µm range**

We used the two thin sections and the thick section to obtain reflectance spectra in the UV-NIR range (from 0.3 to 2.2 µm). The spectrometer was a Shimadzu UV3600, the same we used in previous studies, where the procedure to use this technique has already been explained in detail (Trigo-Rodríguez et al. 2011, 2014; Moyano-Cambero et al. 2013). It allows us to obtain the absolute reflectance of a meteorite section, which can be compared to the remote spectra of asteroids (Trigo-Rodríguez et al. 2014). This spectrometer uses an Integrating Sphere (ISR) and a $BaSO_4$ substrate to create a standard baseline for calibration, which provides close to a 100% reflectance signal better than 1 σ in the 0.3 to 2.2 µm range.

The spectra obtained with this specific spectrometer always show baseline noise between ~0.8 and 0.9 µm, two instrumental peaks (one from 1.4 to 1.6 µm and the other from 1.9 to 2.2 µm), and become too noisy to be reliable after around 2.1 µm, due to humidity, carbon dioxide and system hardware. Therefore, we had to apply some corrections to avoid showing unreliable data (Fig. 2, A). As can be seen, the region between 0.8 and 0.9 µm does not include usable data, and therefore was deleted. The peaks at the regions from 1.4 to 1.6 µm and from 1.9 to 2.2 correspond to an unsuccessful instrumental correction applied by the software to remove the presence of the $BaSO_4$ substrate from the final spectra. We applied a new correction to the spectra in order to get rid of those peaks, but the final result in those regions became slightly noisy, which could hide some faint feature of interest. Finally, to avoid the loss of uniformity above 2.0 µm, we only include data until that wavelength. The scanned area corresponds to a 2x6 mm$^2$ area, which is below the size of the sample and therefore avoids any contribution from the epoxy or the glass in which the section is mounted. Also, comparing the spectra of both a thick and a thin version of the same sample we saw that the contribution of the glass is very small, if any at all (Fig. 2, B).

We noticed that there are obvious differences between our PCA 91467 spectra in this range and the spectrum shown by Cloutis et al. (2012b) (Fig. 3), which nowadays is the only CH3 spectrum in the RELAB (Reflectance Experiment Laboratory) catalogue. It was acquired from a <75 µm grain size powder at the NASA RELAB facility, in bidirectional reflectance mode with a source and phase angle of 30º, and a 5 nm resolution. Both the 0.9 and the 1.9 µm bands in the RELAB spectrum, which are due to the presence almost pure enstatite (Fe-poor pyroxene)(Klima et al. 2007), are hidden in our spectra by the baseline noise between 0.8 and 0.9, and by the noise produced after correcting the instrumental peak from 1.9 to 2.2, respectively. However, in this particular spectrum those bands are quite weak, and according to Cloutis et al. (2012b) they are deepened by the effect of terrestrial weathering. Therefore we can expect those peaks in the parent body to be even fainter.

The most notable difference between our spectra and the RELAB spectrum is the absence of the deep steep below 0.6 µm in our spectra. The presence of this strong ultraviolet absorption can be produced as a result of terrestrial weathering (Cloutis et al. 2012b). Although we obtained our spectra from regions where the terrestrial alteration is not significant (Trigo-Rodríguez et al. 2014), some minor terrestrial weathering in the selected areas is plausible. A strong heating can make this feature disappear (Hiroi et al. 1993), but this is not consistent with a mostly unaltered CH3 chondrite (Bischoff et al. 1994). Hendrix and Vilas (2006) provided another explanation for the disappearance of this feature, showing that space weathering (represented by addition of nanophase iron), resulted on a decrease of slope in this spectral region, but again this is hardly the case as the effects of space weathering are rarely seen on meteorites. However, those authors also explained how minerals with high iron content produce a similar effect (Hendrix and Vilas, 2006). Indeed, in NiFe metals a higher content in Fe imply flattening between 0.4 and 0.6 µm, while higher Ni provides reddening of the spectra between 0.35 and 1.2 µm (Cloutis et al. 2010). As we work with cut sections, the metal grains in our samples are polished, and therefore their metal content becomes more 'visible' from a spectroscopic point of view than in the powder used in RELAB. Indeed, our spectra seem to be dominated by spectrally featureless and red-sloped Fe-rich metal phases (Gaffey et al. 2002; Cloutis et al. 2010) and not by the presence of enstatite like in the RELAB spectrum (Klima et al. 2007; Cloutis et al. 2012b). As a consequence the features in our spectra and the deep steep below 0.6 µm become much weaker, in a very similar way as what happens in space weathered asteroids (Hendrix and Vilas, 2006), which makes our samples good candidates for meteorite-asteroid comparison. Other differences between the spectra obtained from thin sections and powders also could be considered here, but that would imply a much more extended study far away from the purpose of this paper. However, the strongest variations due to grain size are on overall slope (mostly after ~0.6 µm), and reflectance (Johnson and Fanale, 1973), and we see that those particular features are very similar between our spectra and the RELAB spectrum. Therefore, we can consider our spectra complementary to the spectra obtained from a <75 µm powder of the same meteorite (Cloutis et al. 2012b).

**Spectroscopy in the 2.5 to 30.0 µm (4000 to 333 cm$^{-1}$) range**

With an agate mortar we ground into powder the chip from PCA 91467, and studied it with a Fourier Transform Infrared (FT-IR) spectrometer. This kind of device is mainly used to analyze molecular structures, which allows determining the presence of organic molecules and hydrous components. It is also useful to detect silicate minerals, reasons why it is widely used in mineralogy and geology. In our specific case it is equipped with a Smart Orbit Attenuated Total Reflectance (ATR) accessory, equipment that provides an internal reflectance high resolution spectrum whose peaks have the same positions but different relative intensities than an equivalent absorption IR spectrum (Chemtob and Glotch, 2007). It uses a diamond-based detector, which has a wide spectral range plus a good depth of penetration, and is inert, particularly useful when working with meteorites. It has a spectral resolution of ~1 cm$^{-1}$ and the data sampling is every ~0.5 cm$^{-1}$. Its spectral window goes from 2.5 to 45.5 µm (4000 to 200 cm$^{-1}$, approximately), but we discard the data above 30 µm (~333 cm$^{-1}$), which are much noisier or scattered (Trigo-Rodríguez et al. 2012). For a more detailed explanation of the procedure see previous studies (Trigo-Rodríguez et al. 2014).

**Comparison with data from 21 Lutetia**

To directly compare the spectra obtained of meteorite PCA 91467 with information from the asteroid 21 Lutetia in the 0.3 to 2.2 µm range, we used a Lutetia spectrum taken from the Visible and InfraRed Thermal Imaging Spectrometer (VIRTIS) onboard Rosetta in the near-infrared (0.5 to 5 µm) range (Belskaya et al. 2010; Coradini et al. 2011; Sierks et al. 2011). It has to be taken into account that this spectrum includes some undesired instrumental features mostly in the visible (~0.4 to 0.7 µm) region (Coradini et al. 2011). This is why we also used a Lutetia spectrum from groundbased observations (visible range) obtained with the DOLORES (Device Optimized for the LOw RESolution) instrument in the Telescopio Nazionale Galileo (TNG, more details about this spectrum and the instrument in Belskaya et al. 2010), which was also used by Sierks et al. (2011) (Fig. S6 online material) to compare the OSIRIS Narrow Angle Camera (NAC) and Wide Angle Camera (WAC) spectrophotometry taken in flight during the Lutetia fly-by. In Fig. 4 we compare our PCA 91467 spectra with the mentioned spectra from VIRTIS and DOLORES. To properly compare their shape and features, we normalized them to 1 at 0.55 µm. Band center positions were mostly decided using visual criteria, as only an approximate determination is required for our purposes.

For the comparison in the 2.5 to 25.0 µm (4000 to 400 cm$^{-1}$) range, we used the VIRTIS spectrum between 2.5 and 4 µm (4000 to 2500 cm$^{-1}$), plus information obtained from Lutetia with the Infrared Spectrograph (IRS) of the Spitzer Space Telescope (SST), which covers the wavelength range from 5.2 to 38.0 µm (1923 to 263 cm$^{-1}$, approximately). The IRS covered the whole rotational period of the asteroid with a total of 14 full wavelength spectra, which were averaged to obtain the mean emissivity (or emittance) spectrum we used here (see Barucci et al. 2008 and Lamy et al. 2010 for more details). Note that in the IRS data, the region between 13.2 and 15 µm suffers from an excess emission, and therefore features there should be considered carefully (Vernazza et al. 2011). An emissivity spectrum depends in the radiant flux emitted, and its intensity pattern differs from what can be seen in an ATR spectrum as the one we obtained for this study. The position of the peaks is apparently consistent between these two types of IR spectroscopy, according to previous studies, to the point of direct comparison (see, e.g., Morlok et al. 2014). However, we prefer to be conservative and consider that there are too many differences between these techniques to directly compare their shape, so we use these spectra separately. In the first place, we analyzed some of the features present in the ATR spectrum (Fig. 5). Separately we studied, in Figure 6, some features present in the VIRTIS spectrum, comparing them with information obtained about metal-rich carbonaceous chondrites in previous studies (Osawa et al. 2005). To complement that information, we compared the IRS spectrum with spectra of several meteorites from the ASTER (Advanced Spaceborn Thermal Emission and Reflection Radiometer) (Baldridge et al. 2009) and RELAB spectral catalogues, in a similar way as done before by other authors (Nedelcu et al. 2007; Barucci et al. 2008; Lazzarin et al. 2009). The goal was to see the overall resemblances and differences between them (Fig. 7). There is only one PCA 91467 spectrum in the RELAB catalogue, and none in ASTER, so we chose that spectrum intentionally for the comparison. The other meteorite spectra were selected automatically. To do so, we performed a $\chi^2$ test between the Lutetia spectrum and all the meteorite spectra on this two catalogues, and selected the ones showing a higher correlation with the Lutetia spectrum. They were all normalized and shifted in order to simplify the visual comparison between them. All the spectra selected from the catalogues where obtained from powders with grain sizes < 75 µm, except for the KLE 98300 EH3 spectrum which comes from a sample with a grain size < 5 µm. With respect to the techniques, the spectra from RELAB are biconical reflectance spectra collected with an Off-Axis FT-IR Thermo Nicolet Nexus 870 with PIKE AutoDiff, while the spectra from ASTER are bidirectional reflectance spectra obtained with a Nicolet 520FT-IR spectrometer equipped with a Labsphere integrating sphere (Baldrige et al. 2009).

## RESULTS AND DISCUSSION

### Results in the 0.3 to 2.2 µm region

The UV-NIR spectra from Lutetia are usually featureless, except from some faint absorption bands, which are in general understood as different effects of aqueous alteration (see Barucci et al. 2012, and references therein). Indeed, asteroids larger than 100 km are commonly affected by water related processes, probably because they retained water ice that was melted by internal heating and reached the surface by hydrothermal circulation (Grimm and McSween 1989; Fornasier et al. 2014). Another plausible scenario to explain the presence of features related with hydrous minerals, suggests low velocity collisions with primitive asteroids (Gaffey et al. 2002; Shepard et al. 2008), which would be consistent with the heterogeneity on Lutetia's surface (Rivkin et al. 2011; Barucci et al. 2012).

Several bands have been reported on the different Lutetia spectra so far. First, a narrow band at 0.43 µm, attributed to spin-forbidden crystal field transition related with ferric iron (Hunt and Ashley, 1979), appears in several spectra (Lazzarin et al. 2004; Belsakaya et al. 2010). A broader band between ~0.45 and 0.55 µm, or centered around 0.48 µm, which has also been attributed to the $Fe^{3+}$ spin forbidden (Cloutis et al. 2012b) has been seen in some Lutetia spectra (Lazzarin et al. 2009; Belskaya et al. 2010; Perna et al. 2010). Perna et al. (2010) also identified the $Fe^{3+}$ charge transfer transition in iron oxides band at ~0.6 µm (Feierberg et al. 1985) in spectra obtained from the south pole of Lutetia. Another common band is found between 0.8 and 0.9 µm (Bus and Binzel, 2002; Barucci et al. 2005; Birlan et al. 2006; Belskaya et al. 2010; De León et al. 2011), as a result of $Fe^{3+}$ charge transfer transition in iron oxides (Hunt and Ashley, 1979). Although those features can be found heterogeneously around the asteroid (Barucci et al. 2012), the typical 3 µm band widely attributed to aqueous alteration (Lebofsky, 1978) seems to be more common in the southern hemisphere (Rivkin et al. 2011; Barucci et al. 2012), indicating that aqueous alteration is more extended in those regions of Lutetia (Birlan et al. 2004; Rivkin et al. 2011; Vernazza et al. 2011). It also should be noticed that the usual aqueous alteration band at 0.7 µm attributed to $Fe^{2+} \rightarrow Fe^{3+}$ transfer absorption on phyllosilicates (Vilas and Gaffey, 1989), is not found on the Lutetia spectra (Barucci et al. 2012). Although the 0.7 and the 3 µm bands are often associated, some mechanisms like the transformation of all $Fe^{2+}$ into $Fe^{3+}$ or heating to ~ 800 – 900 K can explain the weakening and even disappearance of the first (Cloutis et al. 2011). However, although the importance of the 3 µm band as an indicator of the fundamental O-H stretching bands of $H_2O$/OH (Lebofsky, 1978), this band is not necessarily associated to aqueous alteration (Gaffey et al. 2002; Rivkin et al. 2002), and can be a consequence of the presence of mafic silicates containing structural OH, troilite, solar-wind implanted H, and other anhydrous origins (see Gaffey et al. 2002, and references therein). Therefore, the lack of the 0.7 µm band could talk against aqueous alteration on Lutetia, but most studies indicate otherwise (see Barucci et al. 2012, and references therein).

The specific asteroid spectra used here, coming from the VIRTIS and DOLORES observations, are just two examples of the space and ground based data obtained observing Lutetia in the ultraviolet to near-infrared range in the last 30 years (Barucci et al. 2012), which cover and represent different regions of the asteroid, showing therefore the variable presence of features described before. The VIRTIS spectrum used here for comparison combines information from about 50% of Lutetia's surface, from the north pole to around the equator (Coradini et al. 2011). The DOLORES spectrum comes from the southern hemisphere (Belskaya et al. 2010). In Fig. 4 we plotted together those two spectra of 21 Lutetia, and our spectra of the CH3 chondrite PCA 91467. They show very similar shape up to 0.9 µm, considering the instrumental origin of the bands at 0.6 µm and 0.9 µm in the VIRTIS spectrum (Coradini et al. 2011). Between 0.9 and 1.4 µm the slope is still considerably close, but after this wavelength the VIRTIS spectrum becomes bluer. In fact, several spectra obtained from this asteroid so far show a relatively high variability in slope in the range between 0.9 and 2.4 µm (Nedelcu et al. 2007).

The 0.9 and 1.9 µm bands in the RELAB PCA 91467 spectrum (Fig. 3) clearly correspond to enstatite with only a minor content on Fe ($Fs_{1-5}$) (Bischoff et al. 1994, Klima et al. 2007; Cloutis et al. 2012b). The red slope in both the RELAB and our spectra between 0.9 and 2.2 µm is consistent with the high content of Fe-Ni metal (Bishchoff et al. 1994; Cloutis et al. 2010; Cloutis et al. 2012b). Our three PCA 91467 spectra show a slight decrease in reflectance between 0.43 and 0.83 µm. We removed the continuum in this region, in order to see if the minor bands described for Lutetia between 0.4 and 0.9 µm are producing this effect (Fig 4, B). The very narrow band at 0.43 µm in the DOLORES spectrum can possibly be distinguished as a very faint feature in our spectra (1 in Fig. 4, B, marked as a dotted line), but it is probably a false band as a consequence of the lower resolution of our instrument compared to DOLORES. The band described between ~0.45 and 0.55 µm for some Lutetia spectra (Lazzarin et al. 2009; Belskaya et al. 2010; Perna et al. 2010) is seen in our three spectra (marked as 2 in Fig. 4, B, between two solid lines). We can also consider the very faint band at 0.61 µm (3 in Fig. 4, B, marked as a dotted line) as the 0.6 µm band described before for Lutetia (Perna et al. 2010). A drop-off starting at 0.8 µm in the PCA 91467 spectra can be seen (4 in Fig. 4, B, marked as a dotted line) which could be the beginning of the band between 0.8 and 0.9 µm described for some Lutetia spectra (Bus and Binzel, 2002; Barucci et al. 2005; Birlan et al. 2006; Belskaya et al. 2010; De León et al. 2011), and the 0.9 enstatite band in CH chondrites (Cloutis et al. 2012b).

Many factors can be the origin of the differences between the spectra obtained from the surface of an asteroid and the polished sections of a meteorite. First, we should consider the nature of Lutetia's surface, and the effect it can have on the spectra obtained from the asteroid. Grain size of this material is an important trait to take into account. It has been shown that variations in grain size produce some degree of reddening in the slope of meteorite spectra (Johnson and Fanale, 1973). For example, decreasing the average grain size of CM chondrites usually results in redder spectra (Johnson and Fanale, 1973; Hiroi et al 1993; Cloutis et al. 2011; among others). The bluer slope seen in the Lutetia spectra could therefore be indicative of a grain size in the surface of the asteroid larger than 75 µm (the grain size of the RELAB sample, which shows an equivalent spectral slope to our samples, as explained before). That would be consistent with the Gundlach and Blum (2013) paper, where they described a mean grain radius of surface regolith for Lutetia of 210 µm (although with a large indetermination between +340 µm and -170 µm). However, polarimetric observations indicate that the regolith surface of Lutetia could be covered, at least partially, by material with a mean grain size of less than 20 µm, due to the accumulation of small particles and fragments coming from impact gardening around the asteroid (Belskaya et al. 2010). If this is correct, a different scenario should be considered. According to Cloutis et al. (1990) the presence of magnetite seems to have a bluing effect (reflectance decreasing toward longer wavelengths) on the spectra of asteroids, the finer the grain size of magnetite the bluer the spectra. This mineral has been found in PCA 91467 and other CH chondrites (see e.g. Sugiura, 2000; Greshake et al. 2002; Chang et al. 2015) and its presence is consistent with the band at 0.48 µm, both in the meteorite and the asteroid (Perna et al. 2010; Cloutis et al. 2012b, and references therein). Although the amount of magnetite on CH chondrites is very small (Greshake et al. 2002), this kind of opaque minerals tend to dominate the spectral signature (Gaffey et al. 2002). Furthermore, a fine grain size can have a deepening effect in absorption bands (Johnson and Fanale, 1973) which could also explain why the 0.43 µm band is far more visible in the DOLORES Lutetia spectrum. Another possibility would require a strong heating mechanism for Lutetia beyond 770 K, as high enough temperatures have a bluing effect on spectral slope (Hiroi et al. 1993). If we consider Lutetia as analogous to the parent body of PCA 91467, with the same or similar evolutionary process, this strong heating scenario must have taken place after the meteorite was released from the asteroid's surface, as CH chondrites show low to no evidence of metamorphism on its parent body (Krot et al. 2002).

**Results in the 2.5 to 25 µm (4000 to 400 cm$^{-1}$) region**

In Fig. 5 we show our ATR spectrum of a chip of the meteorite PCA 91467 (inverted to be more easily compared to emissivity spectra from asteroids), between 3.0 and 25.0 µm. The thermal infrared spectra of meteorites and asteroids in this range are usually studied through the Christiansen peak, the Reststrahlen bands and the Transparency features (Barucci et al. 2008, Vernazza et al. 2011, and others). In any case, the presence of these features in our ATR spectrum cannot easily be compared with the features seen in IR spectra obtained with other techniques (see e.g. Morlok et al. 2006, Lane et al. 2011, for examples about the shape of IR spectra obtained by different spectroscopic techniques). However, a Christiansen peak related to mineralogy and grain size (Salisbury 1993), and that was identified in the IRS spectrum of Lutetia by Barucci et al. (2008) at 9.39 µm or 1065 cm$^{-1}$, can be tentatively identified in the ATR spectrum at 9.46 µm or 1057 cm$^{-1}$, as pointed out in Fig. 5 (indicated by a dotted line labeled as 1). This peak, that always occurs between 8 and 9.5 µm (1250 and 1050 cm$^{-1}$) for silicates (Salisbury 1993), has been identified at around 8.3 µm (1205 cm$^{-1}$) for enstatite chondrites (Izawa et al. 2010), which could imply that Lutetia is less related with these meteorites than with CH3 chondrites. In the 8 to 13 µm (1250 to 770 cm$^{-1}$) region, where the main peaks to compare can be found (Barucci et al. 2008; Morlok et al. 2014), the comparison of specific peaks implies many difficulties. The astronomical data usually have lower signal to noise ratios than laboratory data (see e.g. Barucci et al. 2008). Besides, most asteroids spectra lack specific features probably due to the fine particulate regolith that covers their surface (Lim et al. 2005). Also, the IRS spectrum covers a wide region of the asteroid slightly south of the equator, encompassing a higher mineralogical variability than what can be found on a meteorite sample, and even small changes in composition can affect in the position of most mineralogical peaks.

We also examined the Lutetia IRS spectrum in the beginning of the mid-infrared range, from 2.5 to 4 µm (4000 to 2500 cm$^{-1}$), as in this spectral region several features related with water and organics can be found (see Fig. 6). Comparing with the CC spectra analyzed by Osawa et al. (2005), we saw that the Lutetia IRS spectrum is much more similar to CB and CH chondrites spectra than to any other CC spectra analyzed there. First of all, the peak at 2.709 µm (3692 cm$^{-1}$, indicated with a dotted line labeled as 1 in Fig. 6) attributed to free O-H stretching vibrations of chrysotile (magnesium-rich phyllosilicate of the serpentine group), was only found in one CB and two CH chondrites. Notice that it is different from the peak at 2.714 µm (3685 cm$^{-1}$) found in CI chondrites and corresponding to the serpentine mineral lizardite (Osawa et al. 2005). In the Lutetia spectrum this first peak is preceded by another one at 2.66 µm (3758 cm$^{-1}$) which does not appear in any of the CCs spectra in Osawa et al. (2005), and that we have not identified, yet. A very broad peak centered around 2.94 µm (3400 cm$^{-1}$, dotted line labeled as 2) is also representative of CH chondrites, and similar to the O-H stretching vibrations of CI chondrites (Osawa et al. 2005). In fact, the two peaks at 2.963 and 3.091 µm (3375 and 3235 cm$^{-1}$, respectively, corresponding to the dotted lines labeled as 3 and 4) are typical features related with hydrated phyllosilicates and widespread aqueous alteration (Dyar et al. 2011). Although, as explained before, the peaks closer to 3 µm could also have an anhydrous origin, we base our interpretation on the several previous studies suggesting a certain degree of hydration for Lutetia (see e.g. Barucci et al. 2012, and references therein), and do not take into consideration other minor contributions to these features. The next two dotted lines, labeled as 5 and 6, are at 3.419 µm (2924 cm$^{-1}$) and at 3.504 µm (2854 cm$^{-1}$). In the Lutetia IRS spectrum, peak 5 is much higher than the equivalent identified in Osawa et al. (2005), and looks shifted to a lower wavelength, while the peak number 6 is very weak. Those two peaks have been attributed to symmetric and asymmetric C-H stretching vibrations of aliphatic ($CH_2$ and $CH_3$) organics (Sandford et al. 1991) whether they are the consequence of terrestrial contamination (which is obviously not an option for Lutetia) or they have an extraterrestrial origin.

Finally, in Fig. 7 we can see the comparison between the IRS spectrum of Lutetia and five spectra from the ASTER and RELAB catalogues. The selected spectra correspond to one spectrum of the CV3 CC Allende and one spectrum of the CO3.5 CC Lancé (ASTER), plus a spectrum of the CH3 CC PCA 91467, one from the enstatite chondrite of petrologic type 3 (EH3) KLE 98300, and one spectrum from the ureilite Goalpara (RELAB). As mentioned CVs, COs and ECs have been studied before as possible analogs for the asteroid 21 Lutetia (Lazzarin et al. 2004, 2009; Barucci et al. 2008,

2012; Belskaya et al. 2010; Ockert-Bell et al. 2010; Vernazza et al. 2011; among others), while some ureilites, despite being achondrites (with a high content in carbon), have been proved to have similar spectra in the thermal infrared (Lazzarin et al. 2009). CH chondrites have also been suggested in previous studies (Lazzarin et al. 2009; Coradini et al. 2011). Unfortunately, there is only one spectrum of a CH in these two catalogues, and more data would be necessary for a detailed comparison.

We studied all the main peaks in the ~8 to 12 µm (1250 to 833 cm$^{-1}$) plateau of the IRS spectrum, which arises from the O-Si-O asymmetric stretching vibration (Salisbury, 1993), plus three peaks in the ~13.5 to 15.5 µm (740 to 645 cm$^{-1}$) region (they are highlighted with dotted lines in Fig. 7, while Table 1 shows the relative strength of each peak in every spectrum). It has to be taken into account, however, that some of the selected peaks are not very strong, and therefore could be a product of noise, instead of real peaks. Also, that peaks 8 and 9 in the IRS spectrum of Lutetia are in a region that suffers from an excess emission and should be considered carefully (Vernazza et al. 2011). Based on literature data, Table 2 reports a plausible assignment of them all, attributed to the presence of silicates and phyllosilicates. The most abundant classes of silicates in space are olivine and pyroxene, while phyllosilicates (also called hydrous silicates because they form from hydration of anhydrous silicates) are commonly present in interplanetary dust particles and different CC types, in which a shallow band at 2.75 µm (3636 cm$^{-1}$), associated with OH-bonded silicate vibrations, attests to partial hydration of the silicate component. The main diagnostic band that indicates the presence of cosmic silicates is near 9.8 µm (1020 cm$^{-1}$). Dorschner et al. (1995) identified bands at 9.80 and 9.78 µm (1020 and 1022 cm$^{-1}$) associated with olivine glasses. Thus, band 3 in Fig. 7 (9.87 µm or 1013 cm$^{-1}$) can be due to the presence of amorphous olivines. In contrast to the single peak for olivines, pyroxenes (and in particular enstatite) present two characteristic bands at 9.4 and 10.8 µm (1064 and 926 cm$^{-1}$) according to Zaikowski et al. (1975). These two bands (2 and 5 in Fig. 7) are present in the meteorites samples at 9.39 and 10.84 µm (1065 and 923 cm$^{-1}$). We know that crystalline silicates are present in PCA 91467 in the form of enstatite and small amounts of Mg-rich olivine (Bischoff et al. 1994), and band 6 (centered at 11.33 µm or 883 cm$^{-1}$) matches with the diagnostic band at at 11.3 µm (885 cm$^{-1}$) for crystalline forsterite (Bouwman et al. 2001). In the same way, bands 7, 9 and 10 (at 11.63, 14.49 and 15.25 µm, respectively, or 860, 690 and 656 cm$^{-1}$) can be due to the presence of crystalline forms of enstatite as they match to the featured bands at 11.6, 14.5 and 15.4 µm (862, 690 and 649 cm$^{-1}$) of synthetic crystalline clinoenstatite (Jäger et al. 1998). Band 8, at 13.87 µm (721 cm$^{-1}$) can also be related to crystalline enstatite (Jäger et al. 1998). It is worth mentioning that the bands 3 and 9 (9.87 and 14.49 µm, or 1013 and 690 cm$^{-1}$) can also be contributed by the presence of talc, a phyllosilicate with formula $Mg_3(Si_2O_5)_4(OH)_2$. The presence of this mineral can be understood by the hydration of enstatite and its presence has been detected in natural enstatite samples from peaks at 9.8 – 9.9 µm (1020 – 1010 cm$^{-1}$) and 14.6 – 15.0 µm (685 – 667 cm$^{-1}$) (Jäger et al. 1998). Moreover, other phyllosilicates can also be present since band 4 (10.48 µm or 954 cm$^{-1}$) can be attributed to chlorite, a phyllosilicate with general formula $(Mg, Al, Fe)_{12}(Si, Al)_8O_{20}(OH)_{16}$ and a characteristic band at 10.5 µm (952 cm$^{-1}$), and band 2 (9.39 µm or 1065 cm$^{-1}$) can also result from the presence of serpentine $(Mg_6(Si_2O_5)_2(OH)_8)$, which has a featured band at 9.3 µm (1075 cm$^{-1}$) (Zaikowski et al. 1975). Finally band 1 (8.79 µm or 1138 cm$^{-1}$) can be due to the presence of pure silica $SiO_2$ component, but only if we very tentatively consider a broad band centered at

8.6 µm (1163 cm$^{-1}$) present in the spectral region of a sample of Herbig Ae/Be stars (Bouwman et al. 2001).

Table 1.

Table 2.

In the transparency feature between ~12 and 13.5 µm (833 to 740 cm$^{-1}$) and the region above 16 µm (625 cm$^{-1}$), we consider the general shape but not the peaks, as they are poorly defined probably as a consequence of the low signal to noise ratio (Barucci et el. 2008). The general shape of the five spectra match quite well the Lutetia spectrum, but going into more detail several differences can be seen. First, the slope of the smooth plateau between ~13.5 to 15.5 µm (740 to 645 cm$^{-1}$) is similar in all the spectra, but the CCs match better the general relative intensity of the peaks and deep of the bands, although none of them show exactly the same features as Lutetia. Concerning the Christiansen peak, all the catalogue spectra show it shifted by a maximum of ± 0.25 µm with respect to the IRS spectrum (to the point of being confused or mixed with other peaks). The spectrum that apparently shows the closest transparency feature between ~12 and 13.5 µm (833 to 740 cm$^{-1}$) is the one from PCA 91467. For the peaks in the ~13.5 to 15.5 µm (740 to 645 cm$^{-1}$) region, they are absent in Allende and very faint in the other CCs, while they are quite strong in the spectra of Lutetia, the EC and the ureilite. In the region above 16 µm (625 cm$^{-1}$) Allende shows the more similar slope and general behavior, while PCA 91467 and Lancé have slightly bluer spectra and the spectra from KLE 98300 and Goalpara show much more strong features than the IRS spectrum. Several reasons can explain the differences between these spectra. Besidesspecific variations in mineralogy and composition, the grain size of the samples can affect the general slope and the depth of bands, an effect that seems to be more important in CCs than in ECs (Barucci et al. 2012, and references therein). Other variations can be the result of using different techniques. As a summary, it seems that all of them show some spectral similarities with the asteroid 21 Lutetia at those wavelengths, so none of them can be completely discarded or selected as a proper analogue without a more detailed comparison, at least not from a thermal infrared point of view.

**Discussion**

As described, the interpretation of 21 Lutetia spectra seems to indicate the presence of reducing minerals such as enstatite, and also the presence of water-related features. The simultaneous presence of both elements most likely requires formation in an environment with variable conditions and chemical mixing, such as the middle region of the main asteroid belt (Grimm and McSween, 1993; Hardensen et al. 2011), where Lutetia belongs (Barucci et al. 2005). The apparent dissimilitude between Lutetia and this particular CH chondrite could be envisioned in the context described by Davis et al. (1989), in which large asteroids were collisionally disrupted, losing partially their surface layers in the case of asteroids larger than 100 km. In the case of objects with internal ice, heating from decay of their radioactive elements would melt the ice and produce water mobilization and aqueous alteration on portions of the asteroids, possibly close to the surface (Grimm and McSween, 1989), while the interior could be heated to a degree in which thermal metamorphism would be important (Grimm and McSween, 1993). This scenario assumes that water was present on Lutetia before being heated, but an alternative scenario would be considering that CC-type material (and therefore rich in water and carbon) was accreted to its surface, or maybe a water-rich object impacted Lutetia at a slow speed, thus avoiding the vaporization of water (Gaffey et al. 2002; Shepard et al. 2008; Vernazza et al. 2011). That would explain the abundance of phyllosislicates on CH chondrites and their possible detection in the IRS spectrum of Lutetia, while a large portion the asteroid's surface seems to be dry. PCA 91467 would have been formed, or accreted, in the cold surface of such an asteroid, where the aqueous alteration due to internal heating would be very small, or even absent in some regions. Posterior collisions would gave the surface material the breccia constitution that can be recognized on PCA 91467, with a high proportion of fragmented components (Bischoff et al. 1994; 2006), and increase the formation of phyllosilicates. Indeed, CH chondrites (but also CR and CB chondrites) have been affected to some degree by melting, vaporization, outgassing, condensation and size-sorting in a cloud of impact ejecta (Wasson and Kallemeyn 1990). Later, the parent fragment of PCA 91467 would be ejected from the parent asteroid, which surface would finally be reshaped by posterior collisional processes, uncovering superior layers with materials more affected by aqueous alteration, and deeper thermally metamorphosed regions. Applied to Lutetia, that model would partially explain the heterogeneity of its surface (Rivkin et al. 2011; Barucci et al. 2012).

## CONCLUSIONS

After comparing UV-NIR and IR spectra from the asteroid 21 Lutetia and the CH3 meteorite PCA 91467, we reached the following conclusions:

a) The distinctive slope and features, together with the degree of absolute reflection, allow us to establish a possible relationship between PCA 91467 and asteroids resembling Lutetia. This asteroid has been related to M-type asteroids in the Tholen taxonomy, and to the asteroids belonging to the X-complex in the Bus-DeMeo taxonomy. Consistently, CH chondrites exhibit reflectance and mineralogical properties that suggest that their parent bodies can be found among moderate albedo asteroids with mostly featureless spectra. Also, the peculiar combination of a high content on carbon and metal common in CH chondrites can explain many of the special properties of Lutetia, or at least of some regions of this asteroid. Finally, the inferred density of this asteroid is quite close to the bulk density calculated for CH meteorites.

b) In the UV-NIR comparison we found that the selected spectra from Lutetia and PCA 91467 are reasonably similar, considering the differences between polished sections and the regolith-covered surface expected for Lutetia. The intermediate to high reflectance of PCA 91467 is consistent with the intermediate albedo of Lutetia. The differences are probably a consequence of some posterior evolution of the parent body, and also due to the space weathering affecting its surface.

c) We compared the IRS spectrum of Lutetia with several spectra of different meteorites extracted from the RELAB and ASTER catalogues. We tentatively related the several features identified to mineralogy. Despite all spectra compared in Fig. 7 show a certain degree of correlation with Lutetia, none of them can be established as a perfect analogue for this specific spectrum of the asteroid. This is expected, as this specific IRS spectrum is the mean of 14 different spectra, probably showing a heterogeneous combination of mineralogies. The position of particular features clearly differs between Lutetia and enstatite chondrites, and also from the COs and CVs used here for comparison. Those three constitute the main types of meteorites suggested before as analogues to Lutetia. The peaks analyzed suggest a possible detection of phyllosilicates in the surface of Lutetia.

d) In the specific comparison between the 21 Lutetia spectra and Osawa et al (2005) and RELAB spectra of PCA 91467 in the IR region, we have seen the similarity in the position of organic (in the 2.5 to 4 µm region) and silicate (in the 8 to 16 µm region) associated peaks. Several differences point towards a higher presence of aqueous alteration in Lutetia than in PCA 91467.

e) We propose that PCA 91467 was formed as a breccia on the cold surface of an asteroid similar to Lutetia, or CC-like material was accreted on the surface of a primitive version of this asteroid. After the meteorite was released to space, collisional processes broke the outer layers of the asteroid, revealing more aqueously altered and thermally metamorphosed regions. Therefore, the current heterogeneity on the surface of Lutetia would be showing us different degrees of alteration and mineralogical evolution. We propose that PCA 91467 was formed as a breccia on the cold surface of an asteroid similar to Lutetia. After the meteorite was released to space, collisional processes broke the outer layers of the asteroid, revealing more aqueously altered and thermally metamorphosed regions.

We conclude that PCA 91467, our meteorite samples, and by extension the group of CH chondrites, proceed from a Lutetia-type asteroid, i.e., an asteroid with properties and evolutionary history resembling those of Lutetia. Those meteorites can possibly be considered as analogues for particularly primitive regions on the surface of Lutetia. This could imply that at a certain point on Lutetia's life a reservoir of CH-like material accumulated or formed on its surface, as a result of aggregation and/or impacts. The parent fragments of CH chondrites were saved from differentiation due to the size and general properties of its parent asteroid, which prevented thermal metamorphism due to internal heating, and partially aqueous alteration, to reach the most outer layers of this object. As this is still a tentative scenario, a more careful deciphering of both the absorption bands and slope in the UV-NIR, plus the mineralogical features in the IR, is still needed to establish a proper relationship and understanding the evolutionary context of these and similar objects. In fact, for future studies it would be desirable to use a larger amount of Lutetia spectra and also from several CH3 samples measured in different conditions, in order to establish a stronger connection, or to rule it out.

*Acknowledgments* - Current research was supported by the Spanish Ministry of Science and Innovation (projects: AYA2015-67175-P, CTQ2015-62635-ERC and CTQ2014-60119-P) and CSIC (starting grant #201050I043). J. Ll. is grateful to ICREA Academia program. We thank the Rosetta-VIRTIS team to provide the Lutetia spectrum in digital form. NASA Meteorite Working Group for providing the Antarctic carbonaceous chondrites is also acknowledged. A.R. is indebted to Programa Banco de Santander for a UAB distinguished postdoctoral research contract. This study was done in the frame of a PhD. on Physics at the Autonomous University of Barcelona (UAB) under the direction of JMTR.

# TABLES

Table 1. Subjective comparison between peaks of the spectra in Fig. 7.

| Spectrum | 1<br>8.79[a]<br>1138[b] | 2<br>9.39[a]<br>1065[b] | 3<br>9.87[a]<br>1013[b] | 4<br>10.48[a]<br>954[b] | 5<br>10.84[a]<br>923[b] | 6<br>11.33[a]<br>883[b] | 7<br>11.63[a]<br>860[b] | 8<br>13.87[a]<br>721[b] | 9<br>14.49[a]<br>690[b] | 10<br>15.25[a]<br>656[b] |
|---|---|---|---|---|---|---|---|---|---|---|
| 21 Lutetia | sm | md | md | sm | sm | sm | sm | sm | lg | lg |
| Allende | nd | md[c] | md | sm | sm | nd | lg[c] | nd | nd | nd |
| PCA 91467 | sm | md[c] | nd | nd | sm | nd | sm[c] | sm | sm | sm |
| Lancé | sm | md[c] | md[c] | nd | sm | nd | md[c] | sm | nd | sm |
| KL 98300 | sm | lg[c] | md[c] | nd | lg | lg | sm[c] | lg[c] | md[c] | lg |
| Goalpara | sm | lg[c] | md | nd | md[c] | nd | lg[c] | md[c] | md[c] | md |

[a] Position of the peak in µm.
[b] Position of the peak in cm$^{-1}$.
[c] The peak (if the same) is notably shifted.
*nd* Not detected or shifted in such a way that can be confused with another peak.
*sm* Relatively small peak.
*md* Relatively medium peak.
*lg* Relatively large peak.

Table 2. Assignment of the IR bands presented in Figure 7.

| Band position | | Presence of | Reference |
|---|---|---|---|
| µm | cm$^{-1}$ | | |
| 8.79 | 1138 | $SiO_2$ | (Bouwman *et al.* 2001) |
| 9.39 | 1065 | enstatite | (Zaikowski *et al.* 1975) |
| | | serpentine | (Zaikowski *et al.* 1975) |
| 9.87 | 1013 | olivines | (Dorschner *et al.* 1995) |
| | | talc | (Zaikowski et al. 1975) |
| 10.48 | 954 | chlorite | (Zaikowski *et al.* 1975) |
| 10.84 | 923 | enstatite | (Zaikowski et al. 1975) |
| 11.33 | 883 | crystalline olivines | (Jäger *et al.* 1998) |
| | | crystalline forsterite | (Bouwman et al. 2001) |
| 11.63 | 860 | crystalline enstatite | (Jäger *et al.* 1998) |
| 13.87 | 721 | crystalline enstatite | (Jäger et al. 1998) |
| 14.49 | 690 | crystalline enstatite | (Jäger et al. 1998) |
| | | talc | (Zaikowski et al. 1975) |
| 15.25 | 656 | crystalline enstatite | (Jäger et al. 1998) |

**FIGURES**

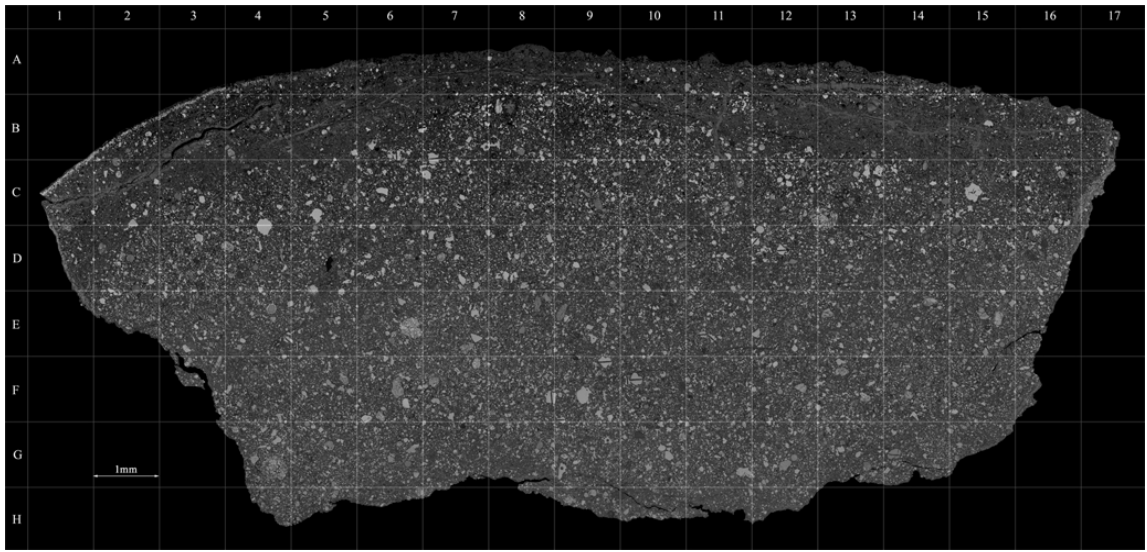

**Fig. 1.** Thin section PCA 91467,25. A thin fusion crust and a 1 mm thick rusty surface that suffered terrestrial weathering can be found at the top. The light grey inclusions are metal grains more abundant below the altered region. The superimposed grid is 1 mm wide.

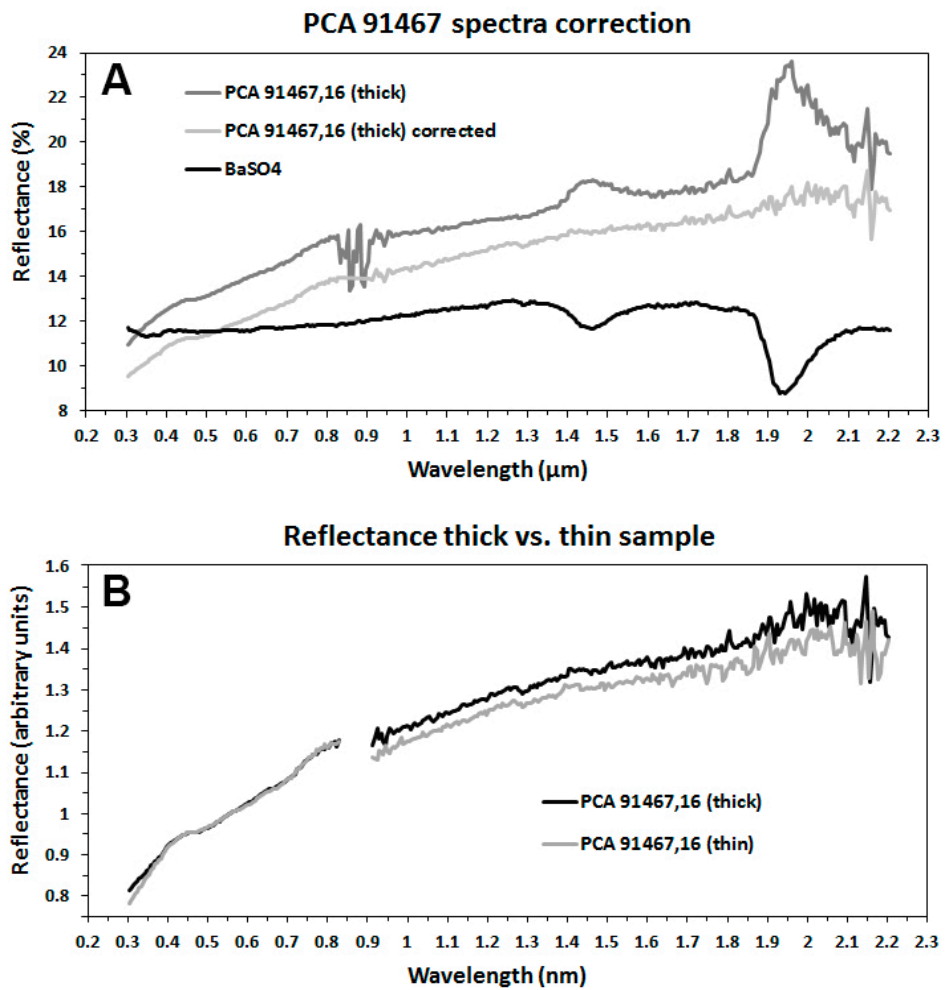

**Fig. 2 (see caption in next page)**

**Fig. 2.** Analysis of the PCA 91467 spectra obtained with the Shimadzu UV 3600 spectrometer between 0.3 and 2.2 µm. In A we show the corrections applied to our spectra. In the region between 0.8 and 0.9 µm instrumental noise dominates the spectra, and therefore we removed these data from the spectra. As can be seen comparing to the BaSO$_4$ spectrum, the peaks between 1.4 and 1.6 and between 1.9 and 2.2 are due to some contribution by the substrate of the spectrometer. Also, above 2.1 µm the signal becomes very noisy and hardly useful. In B we show how the differences between the spectra obtained from a thin and a thick samples are rather small.

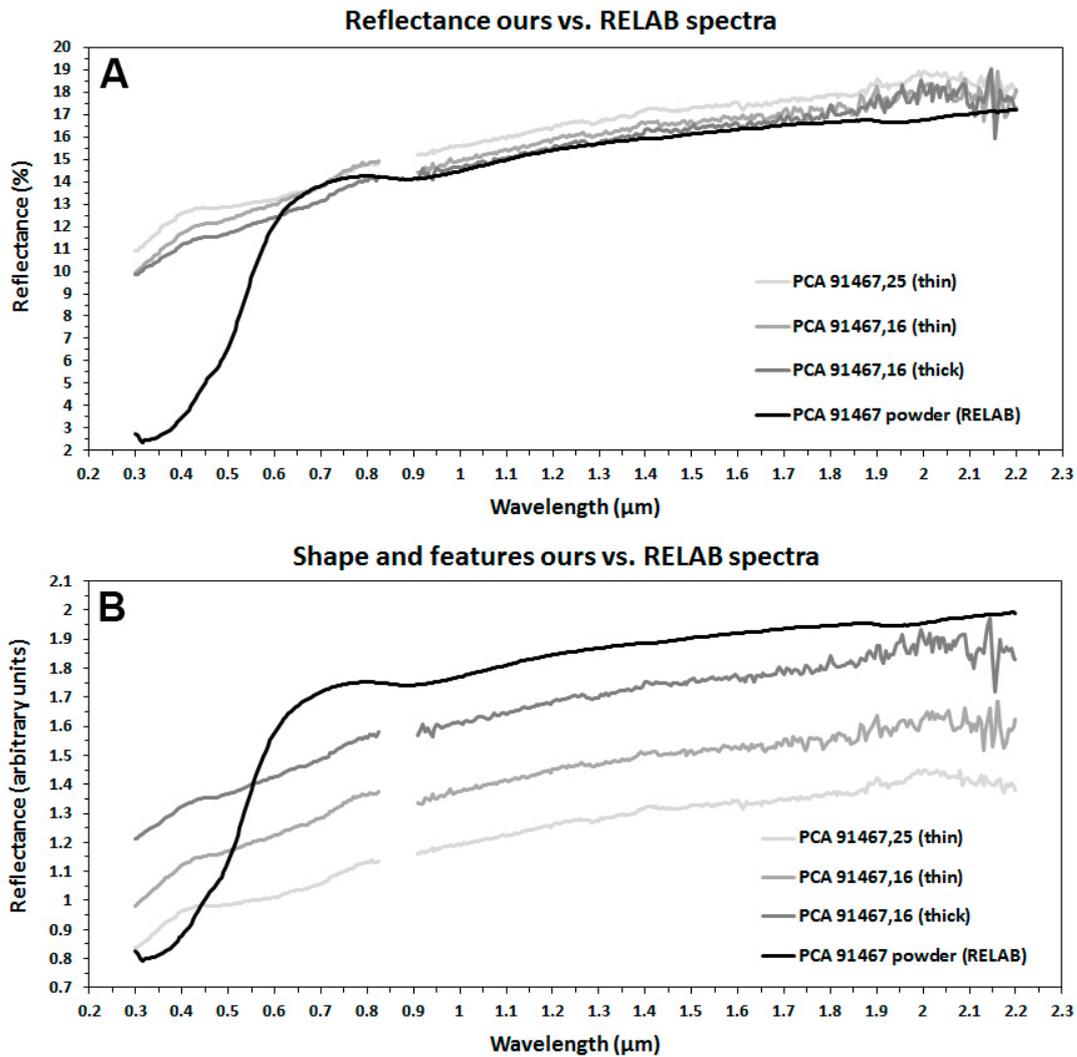

**Fig. 3.** Comparison between our spectra obtained from sections and the RELAB spectrum of the PCA 91467 CH3 chondrite, obtained from a powder. In A we compare the reflectance, showing that, from 0.6 µm onwards, the highest difference between any of our spectra and the RELAB spectrum is smaller than 2%. In B we display the four spectra (normalized and then shifted). They are described in the text.

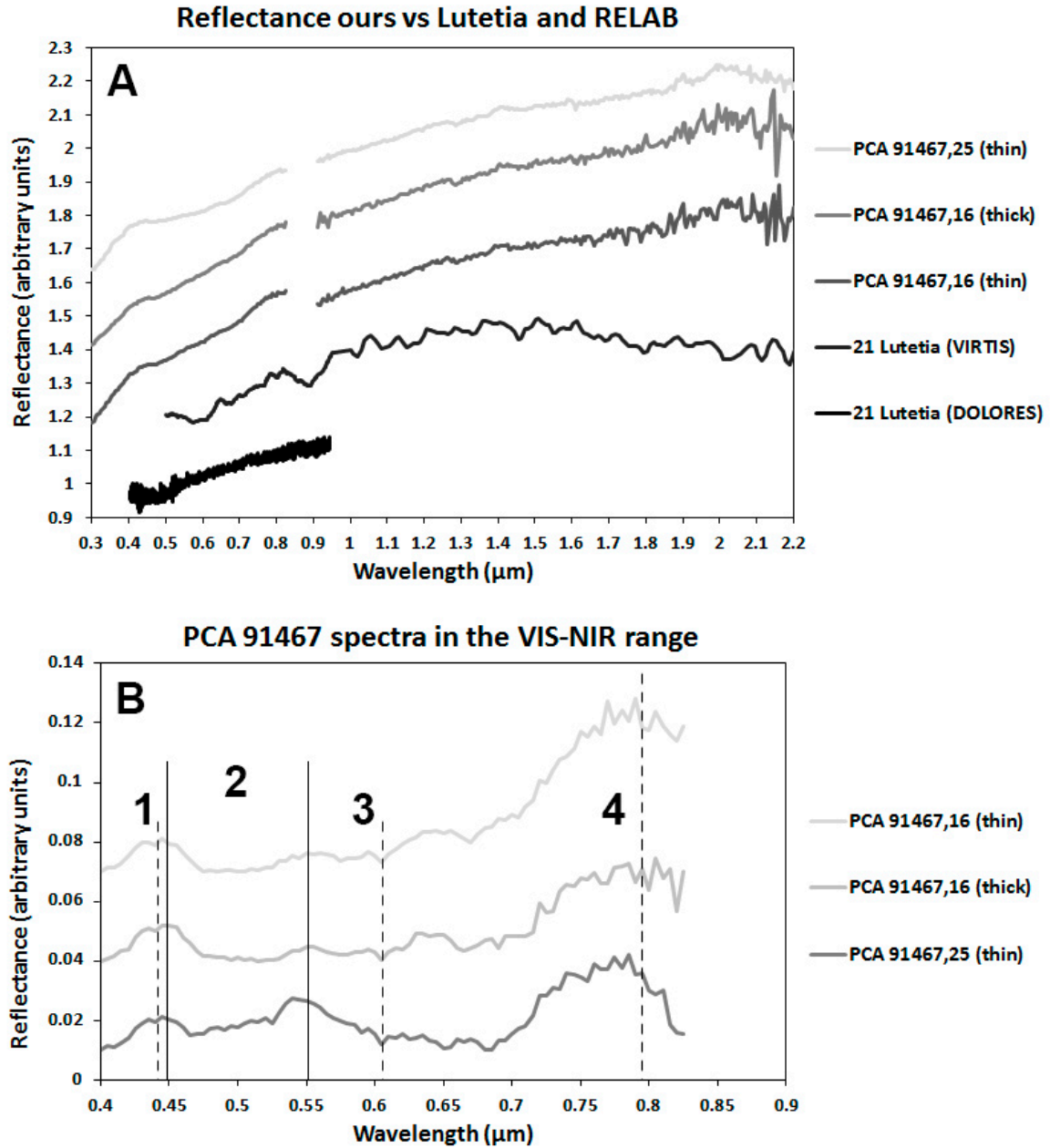

**Fig. 4.** The reflection spectra from the PCA 91467 sections we used compared in A to the Lutetia spectra in the UV-NIR range as measured by VIRTIS and DOLORES. The spectra are normalized to 1 at 0.5 µm, and shifted for visibility purposes. In B we compare the PCA 91467 spectra from our sections to the DOLORES spectrum. The continuum was removed from the 4 of them to properly compare the absorption bands and features. They are also shifted.

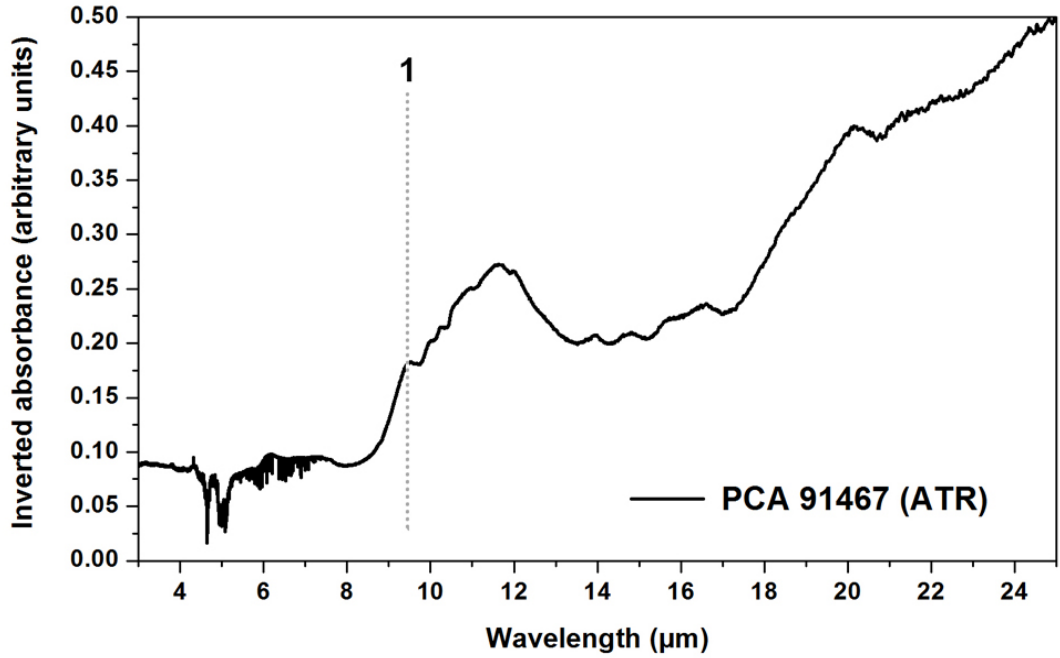

**Fig. 5.** ATR inverted absorbance spectrum of a chip of PCA 91467 (black). The grey vertical dotted line indicates the position of the Christiansen feature.

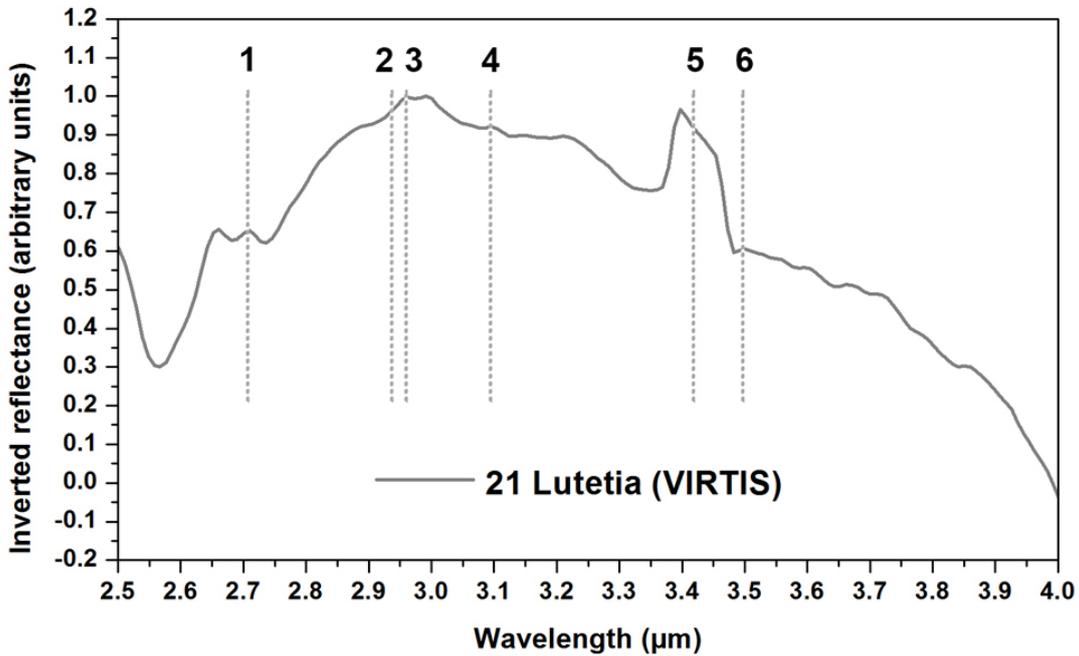

**Fig. 6.** Inverted VIRTIS spectrum of the asteroid 21 Lutetia in the 2.5 to 4.0 µm (and 3800 to 2800 cm$^{-1}$) range. The grey vertical dotted lines indicate the position of several features related to CH chondrites in Osawa et al. (2005), and explained in more detail in the text.

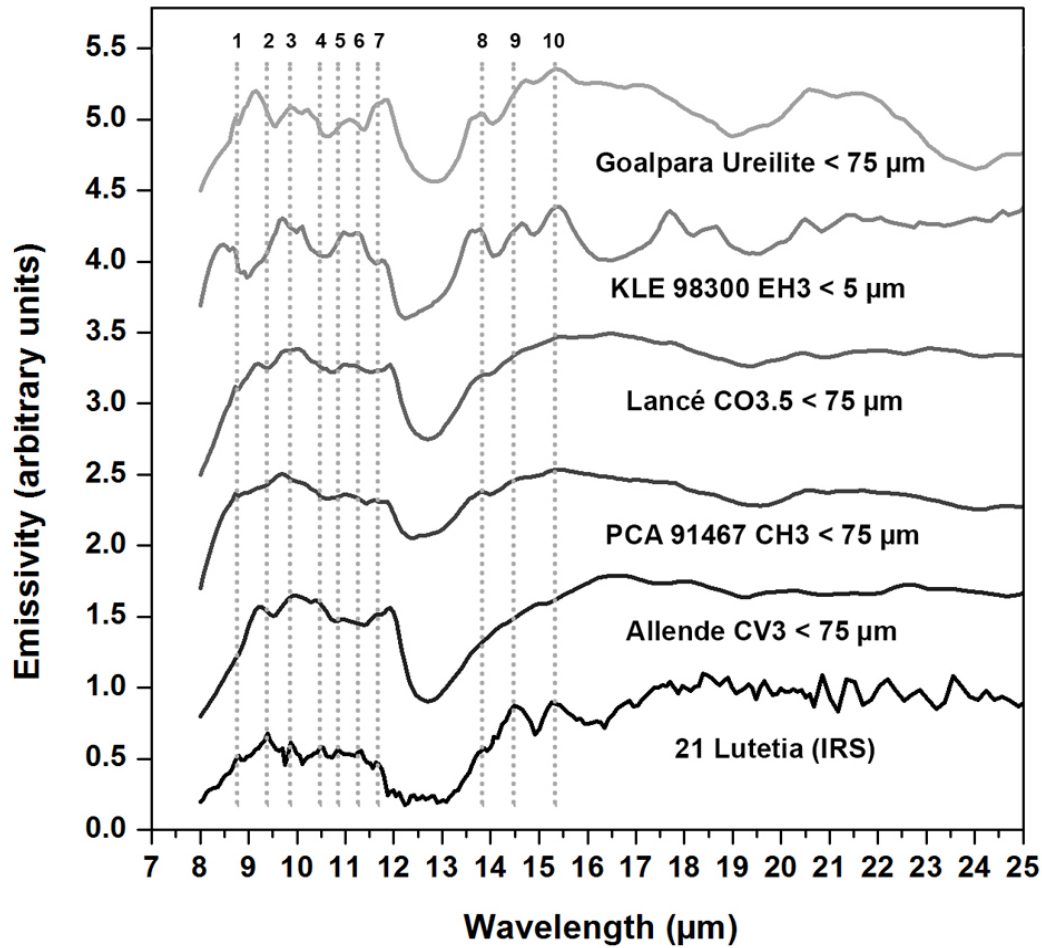

**Fig. 7.** IRS spectrum of the asteroid 21 Lutetia in the 8.0 to 25.0 μm (1250 to 400 cm$^{-1}$) range. The grey vertical dotted lines indicate the position of several peaks in the spectrum of 21 Lutetia (see text and Table for a more detailed explanation). Every spectrum is labeled with its name and grain size. The Allende and Lancé spectra were taken from the ASTER catalogue, while Goalpara, KLE 98300 and PCA 91467 were extracted from RELAB.